\documentclass[published]{JHEP}
\JHEP{11(2001)003}

\usepackage{amsmath,amssymb,graphicx,epsfig,feynmf}

\newcommand{\UU}{\mathop{\rm {}U}}
\newcommand{\SU}{\mathop{\rm SU}}

\newcommand{\be}{\begin{eqnarray}}
\newcommand{\ee}{\end{eqnarray}}
\newcommand{\ra}{\rightarrow}

\newcommand{\cJ}{ {\cal J} }
\newcommand{\cY}{ {\cal Y} }
\newcommand{\cK}{ {\cal K} }
\newcommand{\cI}{ {\cal I} }
\newcommand{\cA}{ {\cal A} }
\newcommand{\half}{ \frac{1}{2} }

\newcommand{\sk}{\sqrt{k}}
\newcommand{\qand}{\quad {\rm and} \quad}
\newcommand{\fk}{\frac{1}{k}}

\newcommand{\fT}{\frac{1}{T}}

\title{Quantum field theory and unification in AdS5}

\author{Lisa Randall and Matthew D. Schwartz\\
	Massachusetts Institute of Technology\\
	Cambridge, MA USA and\\
	Joseph Henry Laboratories, Princeton University\\ 
	Princeton, NJ 08544 USA \\
	E-mail: \email{randall@mit.edu}, 
		\email{matthew@feynman.princeton.edu}}

\abstract{We consider gauge bosons in the bulk of AdS$_5$ in a
two-brane theory that addresses the hierarchy problem.  We demonstrate
that one can do a perturbative calculation above the IR scale
associated with the second brane. We show such a theory can be
consistent with gauge coupling unification at a high scale. We discuss
subtleties in this calculation and show how to regulate consistently
in a bounded AdS$_5$ background.  Our regularization is guided by the
holographic dual of the calculation.}

\received{August 28, 2001}
\accepted{November 2, 2001}


\begin{document}
\begin{fmffile}{fmlong}

\section{Introduction}

In AdS$_5$, the coupling for bulk gauge bosons runs logarithmically,
not as a power law.  For this reason, one can preserve perturbative
unification of couplings. Depending on the cutoff, this can occur at a
high scale. We show that although it is difficult to do this
calculation in a four-dimensional field theory, one can do it in the
full five-dimensional context.  In this paper, we consider the running
of bulk gauge bosons with energy. We discuss several subtleties in the
quantity to be calculated, and in the regularization scheme.  Our
scheme is based on consistency with the holographic correspondence. We
will find that generically, as in the standard model, the couplings
almost unify, if the $\UU(1)$ is normalized consistently with a GUT
group as in $\SU(5)$ \cite{georgiglashow}, as might be the case under
various assumptions for the fundamental physics.  For specific choices
of cutoff and number of scalar multiplets, there is good agreement
with the measured couplings and the assumption of high scale
unification.

This addresses one of the apparent weaknesses of the warped
extra-dimensional theory that addresses the hierarchy problem, that it
appears that one must abandon unification of couplings.  This problem
has been addressed in the context of large extra dimensions in
\cite{dienes1} and \cite{nima}.  However, even if these mechanisms
were to work, one would never have high scale unification. In this
paper, we show that the warped scenario that naturally generates the
hierarchy \cite{rs1} also naturally accommodates unification, due to
the logarithmic running of the couplings. The scale at which
unification occurs depends strongly on the cutoff however; fortunately
this is connected to a potentially observable quantity, the number of
KK modes.

The calculation we do is similar in some respects to that in
\cite{pom}, in which Pomarol considered bulk gauge boson running but
with the assumption that the quarks and leptons live on the Planck
brane. He made the nice observation that unification can occur in this
scenario.  Here, we consider theories that address the hierarchy
through the warped geometry. We also assme the existence of a Higgs on
the TeV brane to generate weak scale symmetry breaking. Another
essential difference in our calculation is in the regularization
scheme.  We will discuss problems with the effective theory
calculation and Pauli Villars regularization. In fact, we will argue
that a specific regularization scheme is dictated by the holographic
correspondence to four-dimensional field theory.

Our analysis also differs from that suggested in
\cite{phenomandhol}. As we will discuss, the particular quantity we
are interested in, the running of the coupling of the massless
four-dimensional mode, can only be calculated explicitly from the
holographic perspective in a theory in which the TeV brane is
explicit. The large tree level logarithmic running does not apply to
the running of the zero mode coupling.

The organization of this paper is as follows: first, we introduce the
theory and review some of the results of the effective 4D picture. We
explore some difficulties with the KK picture for the high scale
calculation required for running the couplings.  In section
\ref{sec5D}, we derive the position/momentum space propagators in the
$R_\xi$ gauges.  We then present some of the Feynman rules and in
section \ref{seclimits} we explore the position/momentum space Green's
functions in various limits.  We introduce and motivate our
regularization scheme in section \ref{secloops}. Through some toy
calculations we show that it is necessary to modify the boundary
conditions and renormalize the Green's functions for energies greater
than $T$ in addition to introducing a position dependent cutoff.  We
then set up an illustrative calculaton, the contribution of a scalar
field to the vacuum energy.  In section \ref{secbeta}, we calculate
the 1-loop $\beta$-functions.  We use the background field method,
where the Ward identities are manifest. Finally, section \ref{secuni}
explores unification in two specific examples. We conclude that
coupling constant unification is entirely feasible in a warped 2-brane
model.

\section{Setup}
As in \cite{rs1}, we postulate the presence of a fifth dimension, and
an anti-deSitter space metric:
\be
ds_5^{2}=\frac{1}{k^2 z^2}(dt^2 - dx^{2} - dz^{2})\,. 
\label{rsmetric}
\ee
We will generally keep the $z$-dependence explicit and contract 4D
fields with $\eta^{\mu\nu}$.  We include two branes: the Planck brane
at $z=1/k$ and the TeV brane at $z = 1/T$.  $T$ is related to the size
of the extra dimension $R$ by $T = k \exp(-kR)$, and defines the
energy scale on the TeV brane. If we take $T \approx$ TeV, we can
naturally explain the weak scale in the standard model.  The fifth
dimension can be integrated out to get an effective four-dimensional
theory, valid at energies below $T$.  The effective 4D Planck scale
$M_{Pl}$ is given by
\be
M_{Pl}^2 = \frac{M^3}{k} \left(1-\frac{T^2}{k^2}\right).
\ee
It is generally assumed that $T \ll k$ and $k \approx M \approx
M_{Pl}$.

Now we put gauge bosons in the bulk. The first question is what is the
quantity we wish to calculate. We know we cannot run the coupling on
the TeV brane above the strong gravity scale, which is roughly
TeV. Furthermore, we are ultimately looking at the zero mode, the only
light mode in the theory at energies of order TeV or below. This is
important; it means the logarithmic running of the coupling considered
in \cite{phenomandhol,eva} does not apply; it is a result of the sum
of all the excited gauge modes, all but one of which are heavy from a
low-energy perspective.  Therefore, the holographic computation of
large threshold corrections does not apply.  We are ultimately
interested only in the relation of the zero-mode coupling of the
four-dimensional theory to the high energy five-dimensional coupling.

\section{Effective four-dimensional theory} \label{seckk}

It is of interest to consider the four-dimensional effective
theory. As mentioned above, we will abandon this in favor of the full
five-dimensional calculation, but here we discuss the theory and why
it is problematic.

The action for a 5D gauge boson is:
\begin{equation}
S_{5D}\! =\! \int\!\! d^4x dz \sqrt{G}\Biggl[\!-\!\frac{1}{4} F_{MN}F^{MN}\!\! +\! \half
\left(a^2k\delta\left(z\!-\!\fT\right)\!+\! \tilde
a^2k\delta\left(z\!-\!\fk\right)\! +\! m^2k^2\right)\! A^M\! A_M\Biggr].
\label{s5d}
\end{equation}
Here $a$ is a mass term on
the TeV-brane, $\tilde a$ is a mass term on the Planck brane,
and $m$ is the bulk mass. The signs are consistent with our metric
convention \eqref{rsmetric}.\\

To begin with, we take the bulk boson to be massless, and set
$a=m=\tilde a=0$. We expand the 5D bosons in terms of an orthonormal
set of KK modes \cite{chang}--\cite{davoudiasl}:
\be
A_\mu(x,z) &=& \sk \sum_{j=0}  A_\mu^{(j)}(x) \chi_j(z)  
\label{amuexp}\\
A_5(x,z) &=& \sk \sum_{j=1} A_5^{(j)}(x)\frac{1 }{m_j}\partial_z \chi_j(z) \,.
\label{a5exp}
\ee

The expansion of $A_5$ is chosen to diagonalize the couplings between
$A_5$ and $A_\mu$ (see below).  Keep in mind that the mass dimensions
are $[z] = -1$, $[g_{5D}]=-1/2$, $[A_M] = 3/2$, $[A_5^{(n)}] =
[A_\mu^{(n)}] = 1$, and $[\chi_n] = 0$.  The eigenfunctions satisfy:
\be
\partial_z(\frac{1}{z}\partial_z\chi_j)= -\frac{m_j^2}{z} \chi_j,
\quad \int \chi_i(z) \chi_j(z) \frac{dz}{z} = \delta_{ij} 
\label{eomchi}
\ee
and therefore have the form:
\be
\chi_j(z) = z(\cJ_1(m_j z) + \beta_j \cY_1(m_j z))\,.
\ee
We assume that the 5D boson, and all the KK modes, have even parity
under the $Z_2$ orbifold transformation. Consequently their
derivatives must vanish on both boundaries. That is, even parity leads
to Neumann boundary conditions. Similarly, odd parity leads to
Dirichlet boundary conditions, which we will discuss in more detail
later on. This leads to the quantization condition:
\be
\beta_j = - \frac{ \cJ_0({m_j}/{k})}{\cY_0({m_j}/{k})}
= -\frac{\cJ_0({m_j}/{T})}{\cY_0({m_j}/{T})}\,.
\ee
For $m_j \ll k$, $\cY_0({m_j}/{k})$ blows up, and so the masses
are basically determined by the zeros of
$\cJ_0({m_j}/{T})$. Therefore, they are spaced in energy by
approximately $\pi T$. This spacing is quite general: it is
independent of bulk or boundary mass terms, and of the spin of the
bulk field; because $\cJ_\nu$ oscillates with the same period for any
$\nu$, bulk fields will always have excitations of order $T$.

We chose the conditions \eqref{amuexp}, \eqref{a5exp} and
\eqref{eomchi} to normalize the kinetic terms and diagonalize the
couplings in the effective 4D action:
\be
S_{4D} = \int d^4x\left[ -\frac{1}{4} (F_{\mu\nu}^j)^2 + \frac{1}{2}
(\partial_\mu A_5^j)^2 -\half m_j^2 (A_\mu^j)^2 + m_j (\partial_\mu
A_5^j) A^{\mu j}\right]. 
\label{s4d}
\ee
This action is explicitly gauge invariant. Indeed, the 5D gauge invariance
\be
A_M \ra A_M - \frac{1}{g_{5D}} \partial_M \alpha(x,z)
\ee
has its own KK decomposition (we expand $\alpha = \alpha_i \chi_i$):
\be
A_\mu(x,z) &\ra&  \sk A_\mu^i(x) \chi^i(z) - \frac{1}{g_{5D}}\partial_\mu \alpha_i(x) \chi^i(z)
\nonumber\\
&=& \sk \left[A_\mu^i(x) - \frac{1}{\sk g_{5D}}\partial_\mu \alpha_i(x)\right] \chi^i(z) 
\nonumber \\
A_5(x,z) &\ra& \sk A_5^i(x) \frac{1}{ m_i}\partial_z \chi^i(z)
- \frac{1}{g_{5D}} \alpha_i(x) \partial_z \chi^i(z)
\nonumber\\
&=& \sk \left[A_5^i(x) - \frac{m_i}{\sk g_{5D}} \alpha_i(x)\right] \frac{1}{m_i}
\partial_z \chi^i(z)\,.\nonumber
\ee
We can plug this back into the action to see that each mode of the 5D
gauge field has an independent gauge freedom.

At this point, it is standard to set $A_5=0$. We see immediately that
this breaks all but the zero mode of the gauge invariance (there is no
$A_5^{(0)}$ since $\partial_z\chi_0(z)=0$).  All modes of $A_5$ are
eaten by the corresponding excited modes of the gauge boson. This is
the unitary gauge. The Goldstone boson ($A_5$) is eliminated and the
massive gauge boson propagators must take the form:
\be
\langle A_5^j(p) A_5^j(-p) \rangle = \frac{-i}{p^2 - m_j^2}
\left(\eta^{\mu\nu} - \frac{p^\mu p^\nu}{m_j^2}\right).
\ee
Although this makes the 4D action look very simple, it is problematic
for evaluating loop diagrams. When we do the full 5D calculation later
on, we will use the Feynman-'t Hooft gauge, in which $A_5$ is included
as a physical particle.

Because it will be useful for interpreting our results, we
present the mass of the lightest KK modes for states of various spin
and mass. For the general lagrangian \eqref{s5d}, with arbitrary mass parameters, the spectrum of a bulk
gauge boson is determined by:
\begin{eqnarray}
&& \frac{\left(-{\tilde a^2}/{2} + 1 -
\nu\right)\cJ_\nu\left({m_j}/{k}\right) + {m_j}/{k}
\cJ_{\nu-1}\left({m_j}/{k}\right)} {\left(-{\tilde a^2}/{2} +
1 - \nu\right)\cY_\nu\left({m_j}/{k}\right) + {m_j}/{k}
\cY_{\nu-1}\left({m_j}/{k}\right)} = 
\nonumber\\
&& \qquad=\frac{\left({a^2}/{2} + 1
- \nu\right)\cJ_\nu\left({m_j}/{T}\right) + {m_j}/{T}
\cJ_{\nu-1}\left({m_j}/{T}\right)} {\left({a^2}/{2} + 1 -
\nu\right)\cY_\nu\left({m_j}/{T}\right) + {m_j}/{T}
\cY_{\nu-1}\left({m_j}/{T}\right)}\,.
\end{eqnarray}
where $\nu = \sqrt{1+m^2}$. If $m$ or $\tilde a$ is nonzero, the
lowest mass is of order $T$.  It cannot be lowered below $T$ unless
$m$ or $\tilde a$ are of the order $T/k \approx 10^{-13}$.  For the WZ
bosons to pick up a weak scale mass without fine tuning, it must come
from the TeV brane. It is not hard to show that for $m=\tilde a =0$
but $a \ne 0$, the lowest mass is given by \cite{hs}:
\be
m_0 \approx \frac{a}{\sqrt{2kR}} T\,.
\ee
If $T =$ TeV, then we can get the weak scale with $a\approx 0.1$.

To get a feel for how the spectrum depends on various parameters, we
present in table~\ref{tab.1} the exact numerical KK spectrum with
$e^{kR} = k/T = 10^{13}$.
\TABLE[t]{\scriptsize{\begin{tabular}{|c|c|c|c|c|c|c|c|c|c|}
\hline
$k/T$ & massless & massive & massive & massive & massless & massless &
      massive & massless & Dirichlet \\
$=10^{13}$ & vector & vector & vector & vector & Dirichlet & scalar &
      scalar & Dirichlet & scalar \\
& & $m=1$ & $m =5$ & $m^2=-1$ & vector & & $m_s=1$ & scalar &
      $m_s=1$ \\ \hline
$m_0/T$ & 0 & 2.869 & 6.873 & 1.297 & 3.832 & 0 & 4.088 & 5.136 &
5.434 \\ \hline
$m_1/T$ & 2.458 & 6.086 & 10.762 & 4.134 & 7.016 & 3.832 & 7.321 &
8.417 & 8.739 \\ \hline
$m_2/T$ & 5.575 & 9.249 & 14.196 & 7.213 & 10.174 & 7.016 & 10.498 &
 11.620 & 11.953 \\ \hline \end{tabular}}
\caption{The exact numerical KK spectrum with $e^{kR} = k/T = 10^{13}$.\label{tab.1}}}

The lagrangian for a massive scalar is $-1/2 \sqrt{G}( \partial_M
\phi \partial^M\phi - m_s^2 \phi^2)$.

Next, we look the couplings between the KK modes in a non-abelian theory.
\be
S &\supset& \int d^4x \left[-g_{ijk} f^{abc} (\partial_\mu A_\nu^{i
a}) A^{j \mu b} A^{k \nu c} -\frac{g_{ijkl}}{4} (f^{eab} A_\mu^{i a}
A_\nu^{j b})(f^{ecd} A^{k \mu c} A^{l \nu d})\right],\qquad
\ee
where the coupling constants are given by overlap integrals.
\be
g_{ijk} = g_{5D} \sk \int \frac{dz}{z} \chi_i \chi_j \chi_k \quad {\rm
and} \quad g_{ijkl} =g_{5D}^2 k \int \frac{dz}{z} \chi_i \chi_j \chi_k
\chi_l\,. 
\label{gijk}
\ee
In flat space, momentum conservation in the fifth dimension implies
that KK number is conserved. For example, $g_{234}$ would have to
vanish, but $g_{224}$ would not.  In curved space this is not true:
$g_{ijk} \ne 0$ in general.  We can say something for the zero mode,
however.  Since we have included no masses, its profile is constant
and equal to:
\be
\chi_0(z) = \frac{1}{\sqrt{kR}}\,. 
\label{chi0}
\ee
Because the $\chi$'s are orthonormal and $\chi_0$ is constant:
\be
g_{0ij} = \frac{g_{5D}}{\sqrt{R}} \delta_{ij} \qand g_{00ij} =
\frac{g_{5D}^2}{R} \delta_{ij}\,.
\ee
If we set $g_{5D}^2 = g_{4D}^2 R$, then the effective theory with just
the zero mode looks identical to a 4D system. Moreover, the zero mode
couples with equal strength to all the KK modes.

It's easy to get a rough idea of how the coupling would run if we took
the effective theory at face value. That is, we assume all the KK
modes are separate particles, and we use a 4D regularization
scheme. At low energy, below $m_1 \approx T$, only the zero mode can
run around the loops. As the energy is increased to $q$, $q/T$ modes
are visible. The result is power law running, similar to what has been
observed in \cite{dienes1} . This is not the correct result.

One possible improvement, suggested by Pomarol in \cite{pom}, is to
regulate with a Pauli-Villars field with a 5 dimensional mass. This
field will have a KK spectrum roughly matching the KK spectrum of the
gauge boson, except that it will have a heavy mode near its 5D mass
instead of a massless zero mode. Thus all the propagators for the KK
modes will roughly cancel and only the zero mode will contribute to
running. It will turn out that this is superficially similar to the
result we will end up getting from the 5D calculation. However,
Pauli-Villars requires that we take the mass of the regulator to
infinity, in order to decouple the negative norm states. But an
infinitely massive field no longer has a TeV scale KK masses, so it no
longer is effective as a regulator. Moreover, it has no hope of
telling us threshold corrections, as unitary is violated in the regime
where the regulator works.

The root of the problem is that the effective theory breaks down at
about a TeV, and so the KK picture is not trustworthy at the high
energy scales necessary to probe unification.  One can calculate on
the Planck brane, but one still has to deal with bulk gauge
bosons. The holographic calculation would be at strong coupling. A
rigorous perturbative approach is to explore the 5D theory directly.

\section{5D Position/momentum space propagators} 
\label{sec5D}

To study the 5D theory, we will work in position space for the fifth
dimension, but momentum space for the other four.

\subsection{$R_{\xi}$ Gauges} 
\label{secrx}

Before gauge fixing, the quadratic terms in the 5D lagrangian are:
\begin{equation}
L = \frac{1}{2kz}\left[A_\mu( \partial^2 \eta^{\mu\nu}
-z\partial_z(\frac{1}{z}\partial_z) \eta^{\mu\nu} -
\partial^\mu\partial^\nu) A_\nu + 2A_5\partial_z \partial^\mu A_\mu -
A_5 \partial^2 A_5\right].
\end{equation}
We would like to set $A_5=0$ and then choose the Lorentz gauge
$\partial_\mu A^\mu=0$. But these conditions are
incompatible. Instead, we will use the following gauge-fixing
functional:
\be
\Delta L = -\frac{1}{2\xi kz}\left[\partial_\mu A^\mu - \xi
z\partial_z\left(\frac{1}{z} A^5\right)\right]^2.
\ee
This produces:
\begin{eqnarray}
L+ \Delta L &=& 
\frac{1}{2kz}\Biggl[A_\mu\left(\partial^2 \eta^{\mu\nu} -z
\partial_z\left(\frac{1}{z}\partial_z\right) \eta^{\mu\nu} -
\left(1-\frac{1}{\xi}\right)\partial^\mu\partial^\nu\right) A_\nu + 
\nonumber\\&&
	\hphantom{\frac{1}{2kz}\Biggl[}
+A_5 (-\partial^2)
A_5 +\xi A_5\partial_z\left( z\partial_z\left(\frac{1}{z} A_5\right)\right)\Biggr]\,.
\end{eqnarray}

We can then read off the equation that the $A_\mu$ propagator must
satisfy:
\be
\langle A^\mu A^\nu\rangle =
-iG_p(z,z')\left(\eta^{\mu\nu}-\frac{p^\mu p^\nu}{p^2}\right)
-i G_{\frac{p}{\sqrt{\xi}}}(z,z')\left (\frac{p^\mu p^\nu}{p^2}\right),
\ee
where $p_\mu = i\partial_\mu$ is the four-momentum, and
\be
\left[ \partial_z^2 - \frac{1}{z}\partial_z + p^2\right]G_p(z,z') = z
k \delta(z-z')\,. 
\label{eomprop}
\ee 
We define $A_5$'s propagator as $\langle A_5 A_5 \rangle \equiv i
\frac{1}{\xi}G^{1,i}_\frac{p}{\sqrt{\xi}}(z,z')$, where
\be
\left[ \partial_z^2 - \frac{1}{z}\partial_z +\frac{1}{\xi} p^2 +
\frac{1}{z^2}\right] G^{1,i}_\frac{p}{\sqrt{\xi}}(z,z') = z k
\delta(z-z')\,. 
\label{eoma5}
\ee
We choose this notation for the following reason.  If we had included
a bulk mass $\half m^2k^2 A_M A^M\sqrt{G}$ in $L$, \eqref{eomprop}
would have been
\be
\left[ \partial_z^2 - \frac{1}{z}\partial_z + p^2 -
\frac{m^2}{z^2}\right]G^{1,m}_p(z,z') = z k \delta(z-z')\,.
\label{eomb}
\ee
Then we can interpret \eqref{eoma5} as saying that $A_5$ has the
Green's function of a vector boson with bulk mass $m^2 = -1$.  For
simplicity we will continue to write $G_p = G_p^{1,0}$ for the gauge
boson.

We can also work out the ghost lagrangian by varying the gauge fixing functional.
\be
L_{\rm ghost} = \frac{1}{kz}c\left( -\partial_\mu D^\mu - \xi z \partial_z
\left(\frac{1}{z} \partial_z \right) \right) c\,.
\ee
Which makes the ghost propagator:
\be
\langle c c \rangle = i \frac{1}{\xi} G_\frac{p}{\sqrt{\xi}}(z,z')\,.
\ee
Note that the ghosts do not couple to $A_5$ directly, as they would
not couple directly to Goldstone bosons in a conventional
spontaneously broken gauge theory.

If we take $\xi = \infty$, we get the unitary gauge.  The gauge boson
propagator looks like:
\be
\langle A^\mu A^\nu\rangle = -i
G_p(z,z')\left(\eta^{\mu\nu}-\frac{p^\mu p^\nu}{p^2}\right) - i
G_0(z,z')\left (\frac{p^\mu p^\nu}{p^2}\right).
\ee
The first term is the transverse polarization states of all the KK
modes.  We can think of the second term as subtracting off the
longitudinal form of the zero mode, Then the zero mode's contribution
has just the regular $\eta^{\mu\nu}$ tensor structure. In this gauge,
the $A_5$ and ghost propagators are zero. It is easy to imagine how
this gauge would make loop calculations very problematic.

$\xi=0$ is the Lorentz gauge. Here the $A_\mu$ propagator is purely
transverse, and $A_5$ and the ghosts have 4-dimensional propagators:
\be
\langle A^\mu A^\nu\rangle = -i G_p(z,z')\left(\eta^{\mu\nu}-
\frac{p^\mu p^\nu}{p^2}\right) \quad\quad \langle A_5 A_5 \rangle =
\frac{i}{R p^2} \quad\quad \langle c c \rangle = \frac{i}{R p^2}\,.
\ee

Finally, $\xi=1$ is the Feynman-'t Hooft gauge. The propagators are:
\be
\langle A^\mu A^\nu\rangle =-i G_p(z,z')\eta^{\mu\nu} \quad\quad
\langle A_5 A_5 \rangle = i G_p^{1,i}(z,z) \quad\quad \langle c c
\rangle = i G_p(z,z)\,.
\label{ftgauge}
\ee
This is the most intuitive gauge. The $A_5$ supplies the longitudinal
polarizations to the excited modes of $A_\mu$.

\subsection{Solving the Green's functions} 
\label{secprop}

To solve \eqref{eomprop}, we first find the homogeneous
solution. Defining $u\equiv \min(z,z')$ and $v \equiv \max(z,z')$, this
is:
\be
G_p(u,v) = u (A \cJ_1(p u) + B \cY_1(p u) )= v (C \cJ_1(p v) + D
\cY_1(p v) )
\ee
$\cJ$ and $\cY$ are Bessel functions. For positive parity under
the orbifold $Z_2$, we must impose Neumann boundary conditions at both
branes:
\be
\partial_u G_p\left(\frac{1}{k},v\right) = \partial_v
G_p\left(u,\frac{1}{T}\right) = 0\,.
\ee
Finally, matching the two solutions over the delta function leads to
the fully normalized Green's function:
\be
G_p(u,v) = \frac{\pi}{2}\frac{ k u v}{AD-BC} ( A\cJ_1(p u)) + B
\cY_1(p u))(C \cJ_1(p v) + D \cY_1(p v)) ,
\label{gp}
\ee
where
\begin{equation}
\begin{array}{rclcrcl}
A &=& -\cY_0(p/k) &\qand& C &=& -\cY_0(p/T) \\
B &=& \cJ_0(p/k) &\qand& D &=& \cJ_0(p/T)\,.
\end{array}
\end{equation}
If the gauge boson has negative parity under the orbifold $Z_2$, then
it must satisfy Dirichlet boundary conditions at both branes (hence we
will call it a Dirichlet boson). Its Green's function will have the
same form as \eqref{gp} but with:
\begin{equation}
\begin{array}{rclcrcl}
A &=& -\cY_1(p/k) &\qand& C &=& -\cY_1(p/T) \\
B &=& \cJ_1(p/k) &\qand& D &=& \cJ_1(p/T)\,.
\end{array}
\end{equation}

Fields of other spin can be derived analogously. For example, the
Green's function for a massless scalar is:
\begin{equation}
S_p(u,v) =\frac{\pi}{2} \frac{k^3 u^2 v^2}{A_sD_s-B_sC_s} (
A_s\cJ_2(p u)) + B_s \cY_2(p u))(C_s \cJ_2(p v) + D_s \cY_2(p v))
\end{equation}
with
\begin{equation}
\begin{array}{rclcrcl}
A_s &=& -\cY_1(p/k) &\qand& C_s &=& -\cY_1(p/T)\\
B_s &=& \cJ_1(p/k) &\qand& D_s &=& \cJ_1(p/T)
\end{array}
\end{equation}

In general, for scalars ($\sigma=2$), fermions ($\sigma = 1/2$) or
vectors ($\sigma = 1$) and with bulk mass $m$, as in $\eqref{s5d}$,
the Green's functions are:
\begin{eqnarray}
G_p^{\sigma m}(u,v)&=& \frac{\pi}{ 2} \frac{ k^{2\sigma-1} u^\sigma v^\sigma}
{A_{\sigma m}D_{\sigma m}-B_{\sigma m}C_{\sigma b}}\times
\nonumber\\&&
\times(A_{\sigma m}\cJ_\nu(p u)) + B_{\sigma m} \cY_\nu(p
u))(C_{\sigma m} \cJ_\nu(p v) + D_{\sigma m}\cY_\nu(p v)\,,
\end{eqnarray}
where $\nu = \sqrt{\sigma^2+m^2}$ and
\be
A_{\sigma m} &=& -\cY_{\nu - 1}(p/k) + (\nu-\sigma)\frac{k}{p} \cY_{\nu}(p/k) 
\nonumber\\
B_{\sigma m} &=& \cJ_{\nu - 1}(p/k) - (\nu-\sigma)\frac{k}{p} \cJ_{\nu}(p/k) 
\nonumber\\
C_{\sigma m} &=& -\cY_{\nu - 1}(p/T) + (\nu-\sigma)\frac{T}{p} \cY_{\nu}(p/T) 
\nonumber\\
D_{\sigma m} &=& \cJ_{\nu - 1}(p/T) - (\nu-\sigma)\frac{T}{p} \cJ_{\nu}(p/T)\,.
\ee
Similar results for KK decompositions can be found in
\cite{gpom}. Keep in mind that although $A_5$ and the ghosts are
scalars, their propagators involve the spin-1 Green's
function. Intuitively, this is expected because they are necessary for
gauge invariance.

We will eventually have to perform a Wick rotation, so that a
euclidean momentum cutoff can be imposed on all the components of
$p_\mu$. To this end, we will need the Green's functions with $p \ra i
q$. These functions are still real. It is easiest to get them by
re-solving equations like \eqref{eomprop} with $p^2 \ra -q^2$. The
result is:
\begin{eqnarray}
G_q^{\sigma m}(u,v)&=& \frac{ k^{2\sigma-1} u^\sigma v^\sigma}
{A_{\sigma m}D_{\sigma m}-B_{\sigma m}C_{\sigma m}}\times
\nonumber\\&&
\times(A_{\sigma m}\cK_\nu(q u)) + B_{\sigma m} \cI_\nu(q
u))(C_{\sigma m} \cK_\nu(q v) + D_{\sigma m}\cI_\nu(q v)\,,
\end{eqnarray}
where $\nu = \sqrt{\sigma^2+m^2}$ as before and
\be
A_{\sigma m} &=& \cI_{\nu - 1}(q/k) - (\nu-\sigma)\frac{k}{q} \cI_{\nu}(q/k) 
\nonumber\\
B_{\sigma m} &=& \cK_{\nu - 1}(q/k) + (\nu-\sigma)\frac{k}{q} \cK_{\nu}(q/k) 
\nonumber\\
C_{\sigma m} &=& \cI_{\nu - 1}(q/T) - (\nu-\sigma)\frac{T}{q} \cI_{\nu}(q/T) 
\nonumber\\
D_{\sigma m} &=& \cK_{\nu - 1}(q/T) + (\nu-\sigma)\frac{T}{q} \cK_{\nu}(q/T)\,.
\ee

\section{Feynman rules}

It is fairly straightforward to derive the Feynman rules in these
coordinates. External particles are specified by their 4-momentum and
their position in the fifth dimension. The vertices have additional
factors of the metric which can be read off the lagrangian. Both loop
4-momenta and internal positions must be integrated over. The Feynman
rules for a non-abelian gauge theory are (in the Feynman-'t Hooft
gauge):
\be
\parbox{20mm}{
\begin{fmfgraph*}(50,25)
\fmfleft{v1}
\fmfright{v2}
\fmf{photon}{v1,v2}
\fmfv{label=$z$}{v1}
\fmfv{label=$z'$}{v2}
\fmfdot{v1,v2}
\end{fmfgraph*} } \ & = & -i G_p(z,z') \eta^{\mu\nu}\\
\parbox{20mm}{
\begin{fmfgraph*}(50,50)
\fmfbottom{vl,vr}
\fmftop{vt}
\fmf{photon}{vt,v}
\fmf{photon}{vl,v}
\fmf{photon}{v,vr}
\fmfv{label=$z$}{v}
\fmfdot{v}
\end{fmfgraph*} } & = & g_{5d} \frac{1}{kz} f^{abc}
[\eta^{\mu\nu}(k-p)^\rho+ \eta^{\nu\rho}(p-q)^\mu +\eta^{\rho\mu}(q-k)^\nu]\\
\parbox{20mm}{
\begin{fmfgraph*}(50,50)
\fmfbottom{vl,vr}
\fmftop{vtl,vtr}
\fmf{photon}{vtl,v}
\fmf{photon}{vl,v}
\fmf{photon}{v,vr}
\fmf{photon}{v,vtr}
\fmfv{label=$z$}{v}
\fmfdot{v}
\end{fmfgraph*} } & = & - i g_{5d}^2\frac{1}{kz} N^{\mu\nu,\rho\sigma}_{abcd}
\ee
where $N^{\mu\nu,\rho\sigma}_{abcd}$ is the standard 4-boson vertex
tensor and group structure.  Then there are the $A_5$ contributions:
\be
\parbox{20mm}{
\begin{fmfgraph*}(50,25)
\fmfleft{v1}
\fmfright{v2}
\fmf{dbl_dots}{v1,v2}
\fmfv{label=$z$}{v1}
\fmfv{label=$z'$}{v2}
\fmfdot{v1,v2}
\end{fmfgraph*} }\ & = & i G^{1,i}_p\\
\parbox{20mm}{
\begin{fmfgraph*}(50,50)
\fmfbottom{vl,vr}
\fmftop{vt}
\fmf{dbl_dots}{vt,v}
\fmf{photon}{vl,v}
\fmf{photon}{v,vr}
\fmfv{label=$z$}{v}
\fmfdot{v}
\end{fmfgraph*} } & = & g_{5d} \frac{1}{kz} f^{abc} (\partial_z^1 - \partial_z^2)\eta^{\mu\nu} \label{aa5}\\
\parbox{20mm}{
\begin{fmfgraph*}(50,50)
\fmfbottom{vl,vr}
\fmftop{vt}
\fmf{photon}{vt,v}
\fmf{dbl_dots}{vl,v}
\fmf{dbl_dots}{v,vr}
\fmfv{label=$z$}{v}
\fmfdot{v}
\end{fmfgraph*} } & = & g_{5d}\frac{1}{kz} f^{abc} (p^\mu - q^\mu) \label{a55}\\
\parbox{20mm}{
\begin{fmfgraph*}(50,50)
\fmfbottom{vl,vr}
\fmftop{vtl,vtr}
\fmf{photon}{vtl,v}
\fmf{photon}{vl,v}
\fmf{dbl_dots}{v,vr}
\fmf{dbl_dots}{v,vtr}
\fmfv{label=$z$}{v}
\fmfdot{v}
\end{fmfgraph*} } & = &  g_{5d}^2\frac{1}{kz}\eta^{\mu\nu}(f^{eab}f^{ecd}+ f^{ead}f^{ecb})
\ee
The derivatives $\partial_z^1$ and $\partial_z^2$ in \eqref{aa5} are
to be contracted with the gauge boson lines, while $p^\mu$ and $q^\mu$
in \eqref{a55} are the momenta of the $A_5$ lines. There are no 3 or 4
$A_5$ vertices because of the antisymmetry of $f^{abc}$.  The Feynman
rules for other bulk fields can be derived analogously, with due
regard for the factors of metric at the vertices. For example, a
$\phi^4$ vertex would have a factor of $(kz)^{-5}$, while a $\phi^2
A_\mu A^\mu$ vertex would go like $(kz)^{-3}$. Ghosts, which are
scalars, technically come from terms compensating for the gauge
invariance of $F_{MN} F^{MN}$, so they have $(kz)^{-1}$ vertices.

\section{Limits of the Green's functions}
\label{seclimits}

Before we evaluate the quantum effects, we will study the propagator
in various limits. To do this, we find it convenient to work with
euclidean momentum. Recall that the Green's function for the massless
vector boson is:
\begin{equation}
G_q(u,v) = k u v
\frac{ [\cI_0({q}/{k})\cK_1(q u) + \cK_0({q}/{k})\cI_1(q u) ]
[\cI_0({q}/{T}) \cK_1(q v) + \cK_0({q}/{T}) \cI_1(q v)]}
{\cI_0({q}/{k})\cK_0({q}/{T})-\cK_0({q}/{k})\cI_0({q}/{T})}.
\end{equation}
One advantage of this form is that the modified Bessel functions,
$\cI$ and $\cK$, have limits which are exponentials, while the
ordinary Bessel functions oscillate.

The first regime we consider is $q \ll T$:
\be
G_{q \ll T}(u,v) \ra -\frac{1}{R\: q^2}\,. 
\label{lowp}
\ee
This is what we expect; at low energy, only the zero mode of the gauge
boson is accessible. Its profile is constant so $G$ is naturally
independent of $u$ and $v$. The factor of $R$ is absorbed in the
conversion from 5D to 4D couplings.

It is also useful to consider the next term in the small $q$ expansion
of $G_q$ at a point $u$ in the bulk.  This will tell us the size of
the 4-Fermi operator which comes from integrating out the excited KK
modes.
\begin{equation}
G_0(u,u) =-\frac{k}{4T^2 k^2 R^2} +\half u^2\left(k + \frac{1}{R}\right) -
\frac{1}{R} u^2 \log ku +\frac{1}{4k^3 R^2}\,.
\label{guu}
\end{equation}

On the Planck brane and TeV branes respectively, it is:
\be
G_0\left(\fk,\fk\right) &\approx& -\frac{k}{4T^2 k^2 R^2} 
\label{gok} \\
G_0\left(\fT,\fT\right) &\approx& -\frac{k}{2 T^2}\,. 
\label{got}
\ee
The additional $(kR)^2$ suppression on the Planck brane over the TeV
brane can be understood from the KK picture. The light modes have
greater amplitude near their masses, which are near the TeV
brane. While there are the same number of heavier modes, localized
near the Planck brane, these are additionally suppressed by the square
of their larger masses. This has been made quantitative by Davoudiasl
et al.\ in \cite{davoudiasl}.  They calculated the equivalent of $G_0$
using the KK mode propagators:
\be
G_0(z,z) = -k \sum \chi_i(z)^2\frac{1}{m_i^2}
\ee
and showed how this number is constrained by precision tests of the
standard model:
\be
V = -M_W^2 R G_0 < 0.0013 \,.
\ee
Using our formula, this forces $T > 8.9$ TeV if fermions are on the
TeV brane and $T > 197$ GeV if fermions are on the Planck brane (for
$kR \approx 32$).

With mass terms, the Green's function satisfies:
\begin{eqnarray}
&& \left[ \partial_z^2 - \frac{1}{z}\partial_z + p^2 - \frac{1}{k^2z^2}
\left(a^2k\delta(z-\fT)+ \tilde a^2k\delta(z-\fk) + m^2k^2\right)
\right] G^{1,m}_p(z,z') = 
\nonumber\\
&&\qquad\qquad = z k \delta(z-z')\,.
\end{eqnarray}
We can work through the same analysis as in the massless case. We find
that if $m$ or $\tilde a$ is not zero, or if $a$ is very large, then
the propagator is constant at low momentum. For a non-zero bulk mass:
\be
G_0^{1,m}(u,v) = -\frac{kuv} { (ku)^\nu (kv)^\nu } \frac{ (1 + \nu +
(\nu-1)(ku)^{2\nu}) (1 + \nu + (\nu-1) (T v)^{2\nu} )} {2 \nu
(\nu^2-1) }\,.
\ee
This tells us the strength of the 4-Fermi operators generated by
integrating out heavy fields. For example, if we have a unified model
where the $X$ and $Y$ bosons get a bulk mass of order $k$, then on the
Planck and TeV branes ($\nu > 1$):
\be
R G_0^{1,m}\left(\fk,\fk\right) &\approx& -\frac{kR}{k^2(\nu-1)} 
\label{xysup}\\
R G_0^{1,m}\left(\fT.\fT\right) &\approx&- \frac{kR}{T^2(\nu+1)}\,.
\ee
We know that constraints from proton decay force this number to be
smaller than $1/(10^{16} {\rm GeV})^2$.  In particular, we are safe on
the Planck brane if $k > 10^{16}$ GeV for any non-zero bulk mass
$m>{k}/{T}$.  On the TeV brane, however, there is no value of the
bulk mass which will sufficiently suppress proton decay; it is
suppressed by at most $1/T^2$.  We clearly need to prevent this
contribution. We discuss this later in the unification section.

Increasing $q$, we find that for $q \gg T$, but $qu \ll 1$
and $qv \ll 1$:
\be
G_q(u,v) \ra -\frac{k}{q^2\left(\log\left({2 k}/{q}\right) - \gamma\right)}\,,
\label{pvll1}
\ee
where $\gamma \approx 0.577$ is the Euler-Mascheroni constant. This is
valid on the Planck brane at $u=v=1/k$ for $q<k$. In particular, it
confirms results of \cite{phenomandhol,pom} that there is a tree level
running of the coupling with $q$. For $q \gg T$ on the TeV brane,
\be
G_q\left(\fT,\fT\right) \ra  -\frac{k}{q T}\,. 
\label{tbra}
\ee
That the propagator goes as $1/q$ instead of $1/q^2$ in this regime is
evidence of what we noted in the effective theory: there are $q/T$
effectively massless modes which contribute at energy $q$. It is also
evidence that this propagator is not valid for $q > T$ on the TeV
brane.

Next, we look at $q \gg T$ and $qu, qv > 1$. Here the propagator looks like:
\be
G_q(u,v) \ra -\frac{k\sqrt{uv}}{2q}e^{-q(v-u)}\,. 
\label{pvgg1}
\ee
The $1/q$ dependence has the same explanation as \eqref{tbra}.  Note
that the propagator vanishes unless $u$ and $v$ are nearly coincident
in the fifth dimension.  Finally, we can consider very large energy,
$q \gg k$:
\be
G_q(u,v) \ra -\frac{k \sqrt{uv} }{q}
\frac{\cosh(q(u-1/ k))\cosh(q(1/ T -v))}{\sinh(q(1/T -1/k ))}\,.
\ee
Since we are at energies much higher than the curvature scale, $q \gg
k$, we get a result very similar to the propagator in flat space with
the fifth dimension bounded at $1/k$ and $1/T$:
\be
G^{flat}_q(u,v) \ra -\frac{1}{q}
\frac{\cosh(q(u-1/k ))\cosh(q(1/T -v))}{\sinh(q(1/T -1/k ))}\,.
\ee

\section{Regulating 5D loops} 
\label{secloops}

From studying the 5D propagator in the previous section, we have
learned that it cannot be trusted for $qu \gg 1$. This is to be
expected if the physical cutoff is around $k$, since the cutoff on the
momentum integral will scale with position in the fifth dimension. So
we understand that we need a position-dependent cutoff on
four-dimensional momentum.  The obvious way to implement this cutoff
is to integrate up to momentum $q = \Lambda/(kz)$ at a point $z$ in
the bulk.  But we will now show that the Green's function must also be
renormalized. The correct procedure is to recompute the Green's
function at an energy $q$ with boundary conditions from an effective
IR brane at $z =\Lambda/(kq)$, and then perform the integral.

Consider the following diagram, which contributes to the gauge boson self energy:
\vskip 0.3cm
\be
\parbox{25mm}{
\begin{fmfgraph*}(60,60)
\fmfleft{v1}
\fmfright{v2}
\fmf{photon,label=${\bf p}$}{v1,vm}
\fmf{photon,label=${\bf p+q}$,tension=0.5}{vm,vm}
\fmf{photon,label=${\bf p}$}{vm,v2}
\fmfv{label=$u$,label.angle=-90}{vm}
\fmfv{label=$z$}{v1}
\fmfv{label=$z'$}{v2}
\fmfdot{vm}
\end{fmfgraph*} }
\quad  &\propto&
g_{5D}^2\int d^4q
\int \frac{du}{ku} G_p(z,u) G_p(u, z') G_{p+q}(u,u)
\ee
We will eventually be concerned with the correction to the zero mode
propagator, so we set the external momentum $p=0$. The low energy
propagator is given by~\eqref{lowp} which is independent of $u$ and
$v$. Since the tree level potential is proportional to $g_5^2/(Rp^2)$
we make the identification $g_{5D}^2 = g_{4D}^2 R$ as before. Also, we
will assume the Ward identities are still satisfied and pull out a
factor of $q^2/p^2$.  Then the integral reduces to:
\be
\frac{1}{R p^2} g_{4D}^2 \int q dq \int \frac{du}{ku} G_q(u,u)\,.
\ee
The $1/(Rp^2)$ in front corresponds to the tree level propagator we
are modifying.  That factor of $1/R$ gets absorbed when we cap the
ends with a $g_{5D}^2$ in a full S-matrix calculation.

Later on, we will calculate this integral exactly, but for now, we only
which to elucidate the regularization scheme. So for a toy calculation,
we will pretend that there is only a zero mode, and so
$G_p(u,u)$ has the $p \ll T$ form \eqref{lowp} at all energies.
First, suppose we have a flat cutoff, at $q=\Lambda$. We know this is wrong,
but if we just have the zero mode, it should give precisely the 4D result.
Indeed,
\be
\int_\mu^{\Lambda} q dq \int_{1/k}^{1/T} \frac{du}{ku} \frac{1}{q^2 R}
= \log\left(\frac{\Lambda}{\mu}\right)
\ee
which is just what we want. Now, suppose we cut off $u$ at $\Lambda/(kq)$,
with this Green's function. Then we have:
\be
\int_\mu^{\Lambda} q dq \int_{1/k}^{\Lambda/(kq)} \frac{du}{ku}
\frac{1}{q^2 R} = \frac{1}{2kR} \log^2\left(\frac{\Lambda}{\mu}\right)
\approx \frac{1}{2} \log\left(\frac{\Lambda}{\mu}\right),
\ee
where we have taken $\mu = \Lambda T/k$ in the last step. Only half
the contribution of the zero mode shows up because at an energy $q$,
we are only including $\Lambda T/(kq)$ of it.

Now suppose the IR brane were at $u = \Lambda/(kq)$ instead of
$u=1/T$.  Then, at any energy, there would always be an entire zero
mode present.  The Green's function would not have a sharp cutoff, but
would get renormalized with Neumann boundary conditions appropriate to
its energy scale. Of course, the physical brane is still at $u=1/T$,
but the Green's function sees the cutoff as an effective brane. With
this regularization, our integral is:
\be
\int_\mu^{\Lambda} q dq \int_{1/k}^{\Lambda/(kq)} \frac{du}{ku}
\frac{k}{q^2\log\left({\Lambda}/{q}\right) }
=\log\left(\frac{\Lambda}{\mu}\right).
\ee

For further illustration, we can work with the full propagator, instead
of just the zero-mode approximation, using our new
regularization scheme. It is natural to split the integral into two regions,
where the propagator can be well-approximated. We can
use \eqref{pvll1} for small $qu$ and \eqref{pvgg1} for large $qu$. The
small region gives, cutting off at $qu=c$:
\be
\int_\mu^{ck} q dq \int_{1/k}^{c/q} \frac{du}{ku}
\frac{k}{q^2\left(\log\left({2k}/{q}\right) - \gamma\right)} \approx
\log\left(\frac{ck}{\mu}\right).
\ee
This is the contribution of one gauge boson, although it is not exactly the ground state.
The large $qu$ region gives:
\begin{eqnarray}
&& \int_\mu^{c k}q dq \int_{c/q}^{\Lambda/(kq)}\frac{du}{ku}
\frac{ku}{2q} + \int_{c k}^{\Lambda} q dq \int_{1/k}^{\Lambda/(kq)}
\frac{du}{ku} \frac{ku}{2q}=
\nonumber\\&&
\qquad\qquad = \frac{\Lambda - ck}{2k}\log\left(\frac{ck}{\mu}\right) +
\frac{\Lambda}{2k}\log\left(\frac{\Lambda}{ck}\right) - \frac{\Lambda
- ck}{2k}\,.
\end{eqnarray}
This represents, roughly, the additional contribution from the excited
modes.  In total, there is a log piece, similar to the 4D log but
enhanced by a factor of $(\Lambda/k - c)/2$. and a constant piece
proportional to $\Lambda$.  For relatively low values of $\Lambda$,
the logarithm will dominate, and theory looks four-dimensional. The
constant piece contributes to threshold corrections. Later on, when we
calculate the $\beta$-function exactly in section~\ref{secuni}, we
will find similar qualitative understanding to this rough analytic
approximation.

Our regularization scheme applies just as well to higher-loop
diagrams. We can define the Green's function $G_p(u,v)$ as normalized
with a brane at $\Lambda/(pk)$, and zero for $v>\Lambda/(kp)$. This
automatically implements the cutoff, and we don't have to worry about
how to associate the $z$ of a vertex with the momentum of a line.

As a final justification of our regularization scheme, we can look at
a renormalization group interpretation through AdS/CFT
\cite{herman,bala}.  It is well known that scale transformations in
the CFT correspond to translations in $z$. But a scale transformation
in a quantum field theory is implemented by a renormalization group
flow. It follows that integrating out the high-momentum degrees of
freedom in the 4D theory should correspond to integrating out the
small $z$ region of the 5D theory.  Suppose our 5D lagrangian is
defined at some scale $M$. This scale is associated not only with the
explicit couplings in the lagrangian, but also the boundary conditions
with which we define the propagators. The high energy degrees of
freedom are not aware of the region with $z > (\Lambda M/k)^{-1}$,
which includes the TeV brane. Therefore, we are forced to normalize
the propagators with an effective virtual brane at $\Lambda M/k$.  In
this way, we derive the low-energy wilsonian effective action in five
dimensions.  If we follow this procedure down to energies of order
TeV, we can then integrate over the fifth dimension to derive the
four-dimensional effective theory.

\section{Corrections to the radion potential}

As a sample calculation, we compute a two-loop contribution to the
vacuum energy that determines the radion potential \cite{gw,csaki}.
Consider the following diagram contributing to the vacuum energy of a
scalar:
\be
{\parbox{12mm}{
{\begin{fmfgraph*}(25,40)
\fmfleft{i}
\fmfright{o}
\fmf{phantom}{i,v}
\fmf{plain}{v,v}
\fmf{plain,left=90}{v,v}
\fmf{phantom}{v,o}
\fmfv{label=$u$}{v}
\end{fmfgraph*}} }}
\quad  &\propto&
\lambda_{5D} \int_0^{\Lambda T/k} q^3 dq
\int_0^{\Lambda T/k} p^3 dp
\int_{1/k}^{1/T} \frac{du}{(ku)^5} S_q(u,u)S_p(u,u)\,.  \nonumber
\ee
Here $\lambda_{5D}$ is the 5D $\phi^4$ coupling. which has dimensions
of length. It is related to the 4D coupling by $\lambda_{4D} =
\lambda_{5D}(2k^3/(k^2-T^2)) \approx 2 k \lambda_{5D}$.  We have cut
off the momenta at  $q=\Lambda T/k$. The region of integration
with $q$ or $p$ greater than $\Lambda T/k$ has no $T$ dependence, and
hence cannot contribute to stabilizing the extra dimension. At low
energies,
\be
S_q(u,u) \approx -\frac{2 k}{q^2} + k u^2 - \frac{k^3}{4} u^4 -
\frac{\log(k/T)}{k} + {\cal O}(q^2)\,.
\ee
Note the enhanced $u$ dependence of the scalar over the vector
propagator (compare \eqref{guu}).  This expansion, which is quite a
good approximation of the full propagator in the region of integration
we are considering, gives a vacuum energy (ignoring the numerical
constants):
\be
V \approx \lambda_{4D} \lambda^4 T^4  + {\cal O}(k^{-2} T^6\log T)\,.
\ee
This expression has the same $T$-dependence as zero point energy
presented in \cite{goldroth}.

\section{Gauge boson self-energy} 
\label{secbeta}

In order to address the question of unification, we will now look at
how various bulk fields contribute to the 1-loop $\beta$-functions.
There are 6 diagrams that contribute at 1-loop: 2 gauge boson, 1
ghost, and 3 involving $A_5$.  One of these diagrams is particularly
ugly, involving $\partial_z$ acting on the $A_\mu$ propagator.  After
fixing the gauge, evaluating, and summing all these diagrams, we
should get a correction to the gauge boson propagator which is
transverse.  But there is an easier way: the background field
method. The idea is to compute the effective action directly, which at
1-loop only involves evaluating functional determinants. Furthermore,
we have the freedom to choose the external field to be whatever we
like. In particular, we can choose it to be the piece of $A_\mu$ which
is independent of $z$. We will see that the quantum fields for
$A_\mu$, $A_5$ and ghosts effectively decouple. The Ward identities
will be explicitly satisfied, as the diagrams containing each type of
particle will separately produce a transverse correction to the
$A_\mu$ propagator.

\subsection{Background field lagrangian}

First, we separate the gauge field into a constant external piece and a fluctuating
quantum piece:
\be
A_M^a \ra \frac{1}{g_{5D}}\left(A_M^a + \cA_M^a \right).
\ee
We have also renormalized out the coupling. Note that $A_M$ now has
mass dimension $1$ as in four-dimensions.  If we let $D_M$ be the
covariant derivative with respect to $A_M$ only then
\be
F_{MN}^a &\ra& F_{MN}^a + D_M \cA_N^a - D_N \cA_M^a + [\cA_M,\cA_N]^a.
\ee
We also must take our gauge fixing functional to be $A_M$-covariant:
\be
\Delta L = -\frac{1}{2 g^2 \xi k z} G(\cA)^2 = -\frac{1}{2 g^2 \xi
kz}\left[D^\mu \cA_\mu^a - \xi z D_z\left(\frac{1}{z} \cA_5^a\right)\right]^2.
\ee
The lagrangian is then:
\be
L &=& -\frac{1}{4g^2 k z}
(F_{\mu\nu}^a + D_\mu \cA_\nu^a - D_\nu \cA_\mu^a + [\cA_\mu,\cA_\nu]^a)^2+ 
\nonumber\\&&
+ \frac{1}{2g^2 k z} (F_{\mu 5}^a + D_\mu \cA_5^a - D_z\cA_\mu^a +
[\cA_\mu,\cA_5]^a)^2-
\nonumber\\&&
- \frac{1}{2 g^2 \xi kz}\left[D^\mu \cA_\mu^a - \xi z D_z\left(\frac{1}{z}
  \cA_5^a\right)\right]^2.
\ee
At 1-loop, we only need to look at terms quadratic in the quantum fields $\cA_M$.
After an integration by parts, the quadratic lagrangian is:
\be
L_2 &=& -\frac{1}{2g^2kz} \Biggl( \cA_\mu^a \left[ -(D^2)^{ab} \eta^{\mu\nu}
+ (D^\nu D^\mu)^{ab}\eta^{\mu\nu} - \frac{1}{\xi}(D^\mu D^\nu)^{ab}\right]
\cA_\nu^b + 
\nonumber\\&&
	\hphantom{-\frac{1}{2g^2kz} \Biggl(}
+ F^{a\mu\nu} [\cA_\mu, \cA_\nu]^a \Biggr)+
\frac{1}{2g^2kz} \cA_5^a\left(-(D^2)^{ab}\right)\cA_5^b+
\nonumber\\&&
+\frac{\xi}{2g^2kz} \cA_5^aD_z\left(z D_z\left(\frac{1}{z} \cA_5^b\right)\right)
- \frac{1}{g^2 k z}\cA_5^a \left[D_\mu D_z -D_z D_\mu\right]^{ab}
\cA_\mu^b-
\nonumber\\&&
-\frac{1}{2g^2k} A_\mu D_z\left(\frac{1}{z} D_z A_\mu\right) + \frac{1}{g^2kz}
F^{a\mu 5} [\cA_\mu, \cA_5]^a\,. 
\label{bigL}
\ee
We derive the ghost lagrangian from variations of $G(\cA)$:
\be
\frac{\delta G}{\delta \alpha} = D^\mu (D_\mu + f^{abc} \cA_\mu^b) -
\xi z D_z\left[\frac{1}{z}(D_z + f^{abc} \cA_5^b)\right].
\ee
Combining \eqref{bigL} with the ghost lagrangian, using the relation
\be
\cA_M^a [D_R,D_S]^{ab}\cA_N^b = -F_{R S}^a [\cA_M,\cA_N]^a
\ee
the final quadratic lagrangian for the pure non-abelian gauge theory
in $AdS_5$ is:
\be
L_2 &=& -\frac{1}{2g^2kz}\! \Biggl(\!
\cA_\mu^a \left[\! -\!(D^2)^{ab}\eta^{\mu\nu}
\!+\! \left(\!1\!-\!\frac{1}{\xi}\!\right)(D^\mu D^\nu)^{ab}\right] \cA_\nu^b
\!+\! z\cA_\mu^a D_z\left(\frac{1}{z} D_z \cA_\nu^b \right)\eta^{\mu\nu}+
\nonumber\\&&
	\hphantom{-\frac{1}{2g^2kz}\! \Biggl(}\!
+\! 2 F^{a\mu\nu} [\!\cA_\mu, \cA_\nu\!]^a\! \Biggr)
\!+\!\frac{1}{2 g^2kz\!}\left(\! \cA_5^a(-D^2)^{ab} \cA_6^b
\!+\! \xi\cA_5^a D_z\biggl(z D_z\biggl(\frac{1}{z} \cA_5^b\biggr)\!\biggr)\! \right)\!+ 
\nonumber\\&&
+ \frac{1}{g^2kz}\bar c^a\left[-(D^2)^{ab}+ \xi z D_z\left(\frac{1}{z} D_z
\cdot\right)\right] c^b\,.
\ee
Observe that the cross terms between $\cA_\mu$ and $\cA_5$ vanish, as
expected.  At this point, we will specialize to the $\beta$ function
calculation we are interested in. It involves an external zero mode of
$A_\mu$, whose profile is constant in the fifth dimension. We simply
the external field to be the piece of the original field which is
independent of $z$.  We also set $A_5=0$.  as there is no external
$A_5$ component.  This lets us write $\partial_z$ instead of $D_z$.
Then in the Feynman gauge, $\xi =1$, the lagrangian is:
\be
L_2 &=& -\frac{1}{2g^2kz} \left( \cA_\mu^a [ -(D^2)^{ab}\eta^{\mu\nu}]
\cA_\nu^b + z\cA_\mu^a D_z\left(\frac{1}{z} D_z \cA_\nu^b
\right)\eta^{\mu\nu} + 2 F^{a\mu\nu} [\cA_\mu, \cA_\nu]^a \right)+
\nonumber \\&&
+\frac{1}{2 g^2kz}\left( \cA_5^a(-D^2)^{ab} \cA_6^b
+ \cA_5^a D_z\left(z D_z\left(\frac{1}{z} \cA_5^b\right)\right) \right)+ 
\nonumber\\&&
+ \frac{1}{g^2kz}\bar c^a\left[-(D^2)^{ab}+  
z D_z\left(\frac{1}{z} D_z \cdot\right)\right] c^b\,. 
\label{smallL}
\ee
We see immediately that the fields have the propagators we derived
before \eqref{ftgauge}, in the Feynman-'t Hooft gauge.  There are no
cross terms between $A_\mu$ and $A_5$ and we can evaluate the
functional determinant for each field independently. In particular,
$A_5$ is seen as a scalar field transforming in the adjoint
representation of the gauge group, with the same Green's function as a
vector with bulk mass $m^2 = -1$.

\subsection{Functional determinants}

We can now evaluate the functional determinants using standard
textbook techniques \cite{ps}.  There are two diagrams which
contribute, one is spin-dependent (from the $F^{a\mu\nu}$ vertex in
\eqref{smallL}) and vanishes for scalars, and the other is
spin-independent. (The third diagram from the quartic interaction does
not contribute as $d\ra 4$ in dimensional regularization, so we will
ignore it for simplicity.) Both diagrams are independently transverse
in the external momentum. This is evidence that the Ward identity for
the 5D gauge invariance is working.

The spin-dependent diagram gives:
\be
\parbox{30mm}{
\begin{fmfgraph*}(80,80)
\fmfleft{i}
\fmfright{o}
\fmf{photon,label=${\bf p}$}{i,v1}
\fmf{photon,left,label=${\bf p+q}$,tension=0.5}{v1,v2}
\fmf{photon,left,label=${\bf q}$,tension=0.5}{v2,v1}
\fmf{photon,label=${\bf p}$}{v2,o}
\fmfv{label=$u$,label.angle=-0}{v1}
\fmfv{label=$v$,label.angle=-180}{v2}
\fmfv{label=$z$}{i}
\fmfv{label=$z'$}{o}
\fmfdot{v1,v2}
\end{fmfgraph*} }
\quad  &&=
-\frac{1}{2} \int \frac{d^4 p}{(2\pi)^4}A_\mu^a(-p) A_\mu^b(p)
\int\frac{d^4 q}{(2\pi)^4}
\frac{p^2 \eta^{\mu\nu} - p^\mu p^\nu}{q^2(p+q)^2} 4 C_r C(j)\times \nonumber\\
&&
q^2(p+q)^2 2\int_{1/k}^{\Lambda/(kq)}  \frac{du}{ku}
\int_u^{\Lambda/(kq)} \frac{dv}{kv}
G_q(u,v) G_{q+p}(u,v)
\ee
where $C(j)=2$ for vectors and zero for scalars and $C_r$ is Dynkin
index for the appropriate representation.  Since we are interested in
the vacuum polarization in the $p \ra 0$ limit, and the transverse
projector is already manifest, we can simply set $p=0$ in the $uv$
integrals.  Now change variables to $y=qu$ and $z=qv$ and set
$k=1$. Then the second line above becomes
\be
I(\Lambda,q)= q^4 2\int_q^\Lambda \frac{dy}{y} \int_y^{\Lambda}
\frac{dz}{z} G_q\left(\frac{y}{q},\frac{z}{q}\right)^2. 
\label{defI}
\ee
The point of doing this is that the integrand now contains the square
of:
\be
G_q\left(\frac{y}{q}, \frac{z}{q}\right) = \frac{y z}{q^2} \frac{
[\cI_0(q)\cK_1(y) + \cK_0(q)\cI_1(y) ][\cI_0(\Lambda) \cK_1(z) +
\cK_0(\Lambda) \cI_1(z)]}
{\cI_0(q)\cK_0(\Lambda)-\cK_0(q)\cI_0(\Lambda)}\,.
\ee
The $1/q^4$ in $G_q^2$ cancels the $q^4$ prefactor in \eqref{defI},
leaving a dimensionless number multiplying the standard 4D
integral. This is not strictly true, as $I(\Lambda,q)$ still has a
weak dependence on $q$. But quite generally, we can write
$I(\Lambda,q) = I_0(\Lambda) + I_1(\Lambda) \frac{q}{k} + \cdots$, so
for $q \ll k, I = I_0$ is a fine approximation. Anyway, the background
field method at 1-loop cannot give us reliable information about
additional divergences, or threshold corrections. The best we can do
is to use the degree to which $I$ is not constant as a rough measure
of the size of the additional corrections.

The other diagram, which is spin-independent. contributes:
\be
&&-\frac{1}{2} \int \frac{d^4 p}{(2\pi)^4}A_\mu^a(-p) A_\mu^a(p)
\int\frac{d^4 q}{(2\pi)^4}
\frac{(2q+q)^\mu(2q+p)^\nu}{q^2 (p+q)^2}
C(r) d(j) \times 
\nonumber\\&&
\qquad \times q^2 (p+q)^2  2\int_{1/k}^{\Lambda/(kq)}  \frac{du}{ku}
\int_u^{\Lambda/q} \frac{dv}{kv} G_q(u,v) G_{q+p}(u,v)\,,
\ee
where $d(j)$ is the number of spin components.  The second line is
exactly the same integral expression, $I(\Lambda,q)$, as before.
While the tensor structure of the first line is not explicitly
transverse in the external momentum, it is in fact transverse after
the $q$ integral is performed (in dimension regularization as $d\ra
4$).

Each particle will have a different value for $I_0$.  We must replace
$G_q$ with the appropriate propagator (cf.\ section \ref{secprop}) for
each particle and redo the integrals in each case.  We shall call the
result $I_0^{\sigma, m}$, corresponding to $G_p^{\sigma, m}$ from
section \ref{secprop}.  Observe that as with the $R_\xi$ gauges, $A_5$
and the ghosts have kinetic terms corresponding to spin 1, so they
will both have $\sigma=1$. In particular, $I_0$ for ghosts is
identical to $I_0$ for~$A_\mu$.

The reason these diagrams are relevant is that they directly produce
$F^2$ terms in the effective action. Indeed, the Fourier transform of
the quadratic terms in $F^2$~is:
\be
-\frac{1}{4 g^2}\int d^4x \frac{dz}{kz} F_{\mu\nu}F^{\mu\nu} =
-\frac{R}{2 g^2} \int\frac{d^4p}{(2\pi)^4} A_\mu^a(-p) A_\nu^b(p) (p^2
\eta^{\mu\nu} + p^\mu p^\nu)\,.
\ee
Since $A_\mu$ is independent of $z$, we have performed the
$z$-integral explicitly.  So we see that we have calculated a
correction to the dimensionless coupling $g_{5D} R^{-\half}$.
Equivalently, we have calculated the running of $g_{5D}$ itself, once
we absorb the factor of $R$ into the coefficient of the logarithm. The
result for the 1-loop $\beta$-function is:
\be
\beta(g_{5D}) = -\frac{g_{5D}^3}{4\pi^2}
\frac{1}{R}C_2(G)\left(\frac{11}{3}I_0^{1,0}- \frac{1}{6}
I_0^{1,i}\right) + matter\,.
\label{betafn}
\ee
It may appear that we have calculated the running for only the zero
mode of the 5D gauge boson. But gauge invariance implies that there
can only be one 5D coupling, at any energy scale.  So we have in fact
calculated the $\beta$ function of every mode.  If we define $g_{4D}
\equiv g_{5D} R^{-\half}$, then the 4D $\beta$-function is:
\be
\beta(g_{4D}) = -\frac{g_{4D}^3}{4\pi^2} C_2(G)\left(\frac{11}{3}I_0^{1,0}-
\frac{1}{6} I_0^{1,i}\right) + matter\,.
\ee
Note that the sign of the $A_5$ contribution is opposite to that of
the ghosts (which contribute $\frac{1}{3} I_0^{1,0}$ to the above
expression). This is because they have opposite statistics which
changes the sign of the exponent of the functional determinant. It is
easy to understand this result; the ghosts remove the two unphysical
polarizations of $A_\mu$, and $A_5$ adds back one of them.

\subsection{Numerical results for $I(\Lambda,q)$}

\FIGURE[t]{\centerline{\epsfig{file=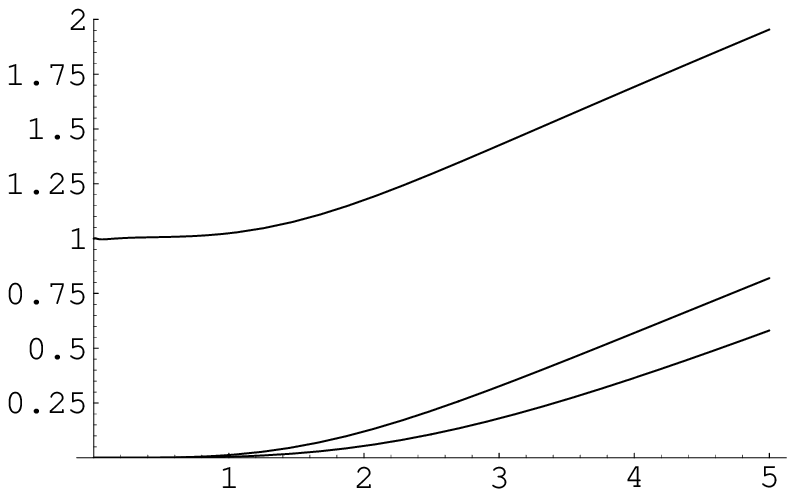,width=.6\textwidth}}%
\caption{$I_0$ as a function of $\Lambda/k$: from top to bottom:
massless vector, massive vector ($m=1$), and Dirichlet
vector.\label{bddm5}}}

The function $I_0(\Lambda)$ is shown in figure \eqref{bddm5}, and
$I(\Lambda,q)$ is shown in figures~\ref{31s} and~\ref{lmdd5}.  Since
we need need to perform a Wick rotation to evaluate the 4D integrals,
we used the euclidean propagators in calculating $I(\Lambda,q)$.  Of
course, the constant piece, $I_0$, is independent of $q\ra i q$, and
we have confirmed this numerically. It turns out that the integrals
converge faster in euclidean space.  We can see from
figure~\ref{bddm5} that $I_0$ is roughly proportional to the number of
KK modes running around the loop.  Because of our boundary conditions,
the effective spacing between the KK modes at an energy $q$ is $\pi q
k/ \Lambda$. For the massless 5D vector, at energy $q$ there is a
massless mode, plus approximately $q / (\pi q k/\Lambda) =
\Lambda/(\pi k)$ other modes visible. This fits roughly with
figure~\ref{bddm5}. The discrepancy is due to the fact that the
spacing is not precisely $\pi$ for the lowest modes, and that the sum
of the higher modes is not completely negligible.  Notice that the
$I_0$'s of massive and Dirichlet cases, which have no massless zero
mode, are about 1 less than $I_0$ for the massless vector.  When
$\Lambda$ is bigger than $k$, and the $beta$-function is
correspondingly higher, the unification scale will be lower.  A theory
of accelerated unification was also considered in ref.~\cite{harvard}
in a different scenario.

\FIGURE[t]{\centerline{\epsfig{file=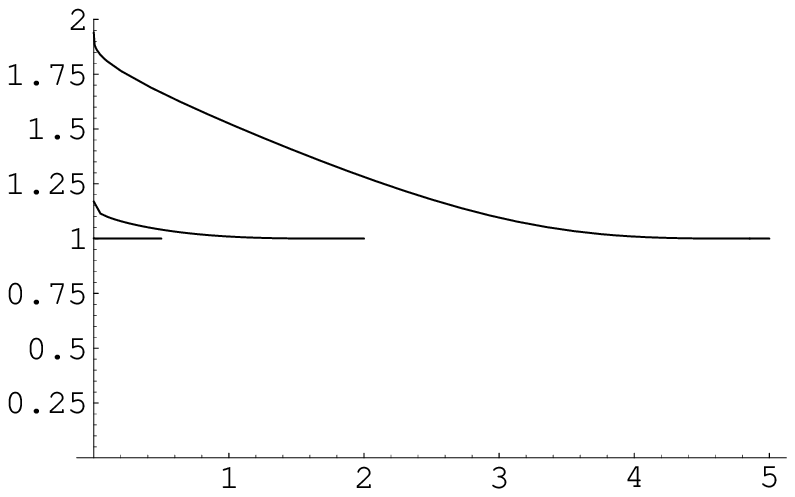,width=.6\textwidth}}%
\caption{$I(\Lambda,q/k)$ as a function of $q/k$ with $\Lambda$
fixed. From top to bottom, $\Lambda = 5k, 2k,$ and $0.5 k$.\label{31s}}}

\FIGURE[t]{\centerline{\epsfig{file=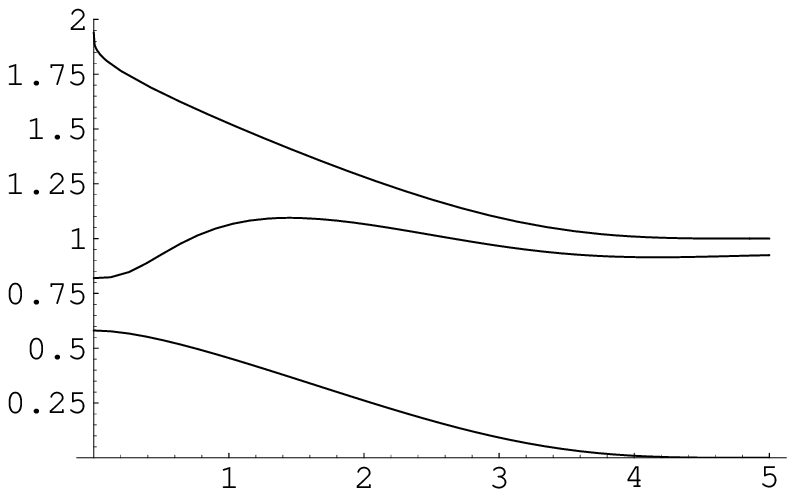,width=.6\textwidth}}%
\caption{$I(\Lambda = 5k,q/k)$. From top to bottom: massless vector,
massive vector (with bulk mass $m=1$), and Dirichlet
vector.\label{lmdd5}}}

\looseness=1Figures~\ref{31s} and~\ref{lmdd5} can be understood similarly. As $q
\ra k$, the branes approach each other.  For the massless case, there
is always one complete mode, the zero mode.  In this limit the theory
looks 4-dimensional. For the massive case, the zero mode exists as
well, and so it approximates the massless case.  With Dirichlet
conditions, the zero mode is eliminated, so the function goes to zero.
Note that if we had left the boundary conditions at $1/T$, this
function would have gone to zero as $q\ra \Lambda$ for any of the cases.
We list the values of $I_0(\Lambda)$ for various cases in table~\ref{tab.2}.

Recall that in the Feynman gauge, ghosts and $A_5$ get the $I_0$ of
vectors and $A_5$ has an effective mass $m=1$.  The numbers in this
table can be predicted approximately from the KK masses in
table~\ref{tab.1}.  We can see that $I_0(\Lambda)$ is approximately
the number of modes with mass less than $\Lambda T/k$. This gives us a
very useful intuition for seeing how changing the field content
affects unification, as we will now illustrate.

\TABLE[t]{\scriptsize{\begin{tabular}{|c|c|c|c|c|c|c|c|c|c|}
\hline
& massless & massive & massive & massive & massless & massless &
massive & massless & Dirichlet \\
$I^{\sigma,m}_0(\Lambda)$ & vector & vector & vector & vector &
Dirichlet & scalar & scalar &Dirichlet& scalar \\
& & $m=1$ & $m =5$ & $m^2=-1$ & vector & & $m=1$ &scalar & $m=1$ \\ 
\hline
$\Lambda =0.5$ & 1.007 & 0.001 & 0.000 & 0.018 & 0.000 & 1.000 & 0.000
& 0.000 & 0.000 \\ \hline
$\Lambda =1 $ & 1.024 & 0.013 & 0.001 & 0.147 & 0.005 & 1.005 & 0.004
& 0.002 & 0.001 \\ \hline
$\Lambda =5 $ & 1.954 & 0.820 & 0.178 & 1.411 & 0.581 & 1.581 & 0.525
& 0.353 & 0.315 \\ \hline
\end{tabular} }
\caption{Values of $I_0(\Lambda)$ for fields of various spin and mass.\label{tab.2}}}

\section{Coupling constant unification} 
\label{secuni}

In order to study coupling constant unification, we need to choose a
particular model.  Because the main motivation of this work is to
solve the hierarchy problem using the warp factor, all weak-scale
masses should generated from Higgs scalars confined to the TeV brane.
We will consider three possible scenarios. The first is that there is
no unified group.  Indeed, the generic prediction of fundamental
theories is only that there should be one coupling constant at high
energy, not that there should be a unified group. We put the 3-2-1
gauge bosons in the bulk, and the Higgs and fermions on the TeV
brane. From the CFT point of view, the TeV brane fields are to be
viewed as condensates. So we can expect there to be a number of bulk
fermions or scalars transforming as electroweak doublets, which
condense to form the Higgs.  In this case, We have to assume there is
at a fundamental level a reason to assume the $\UU(1)$ is normalized
in a way consistent with a GUT model. For this, additional physics
assumptions are necessary.

A second possibility is that there is a unified group, such as
$\SU(5)$.  If the doublet Higgs is part of a larger multiplet, such as
a vector ${\bf 5}$, then the triplet will necessarily have TeV scale
excitations, leading to proton decay.  This is the standard
doublet-triplet problem \cite{lisa23}. One 4D solution is to couple
the triplet to a missing partner, which gives it a large mass and
decouples it from the standard model. However, this solution will not
work with a TeV brane triplet, because its mass can be at most
TeV. Instead, one can for example implement the pseudoGoldstone boson
method, as in \cite{lisa23,cheng}. Briefly, the idea is to postulate a
weakly gauged global symmetry, such as $\SU(6)\times \SU(6)$. This is
broken by an adjoint $\Sigma$ and two fundamentals $H$ and $\bar
H$. The doublet Higgs arises as a pseudoGoldstone boson, and there is
no triplet at all. This sort of condensation also seems likely from
the CFT point of view, where all TeV brane fields are composites.

Dimension 6 operators that violate baryon number pose a potential
problem.  As mentioned before, the $X$ and $Y$ bosons should not
couple on the TeV brane.  This can be done by having the gauge
symmetry not commute with the orbifold transformation \cite{kawa,hall}
so that the $WZ$ have positive parity under $Z_2$, but the $XY$ have
negative parity and have vanishing amplitude on the TeV brane.  An
additional baryon number symmetry should be imposed on the brane to
forbid dangerous operators. Notice that the $\UU(1)$ on the brane
might have a kinetic term, and therefore a coupling, not determined by
unification. If this is the case, one would hope the brane couplings
for $\UU(1)$, $\SU(2)$, and $\SU(3)$ are all big so that a mechanism
such as the one in \cite{neal2} would apply.

A third possibility is that we don't use the TeV brane to generate the
weak scale, as in \cite{pom}.  The hierarchy problem must be solved
some other way, such as using supersymmetry.

Now return to the first scenario, with no unified group. The TeV-brane
particles will contribute to running only up to $q = \Lambda
T/k$. After this, they contribute like the bulk fields which represent
their preonic constituents in the CFT.  There are many possibilities
for what these can be, but for the sake of illustration, we will
assume they are either fermions or scalars which have the 3-2-1
quantum numbers of the standard model Higgs. Then the 1-loop
$\beta$-functions lead to the following running:
\be
\alpha_1^{-1}(M_{GUT}) &=& \alpha_1^{-1}(M_Z) -
\frac{2}{\pi}\left(\frac{n_g}{3} + \frac{3}{5}\frac{n_s}{24} \right)
\log\left( \frac{M_{GUT}}{M_Z} \right)
\label{uni1}\\
\alpha_2^{-1}(M_{GUT}) &=& \alpha_2^{-1}(M_Z) - \frac{2}{\pi}
\left(-\frac{11}{6}I^{1,0}_0(\Lambda) + \frac{1}{12} I_0^{1,i}(\Lambda) +
\frac{n_g}{3} + \frac{n_s}{24} \right) \log\left( \frac{M_{GUT}}{M_Z}
\right)
\nonumber\\
\label{uni2} \\
\alpha_3^{-1}(M_{GUT}) &=& \alpha_3^{-1}(M_Z) - \frac{2}{\pi}
\left(-\frac{11}{4}I^{1,0}_0(\Lambda) + \frac{1}{8} I_0^{1,i}(\Lambda) +
\frac{n_g}{3} \right) \log\left( \frac{M_{GUT}}{M_Z} \right).
\label{uni3}
\ee

The $I_0^{1,i}$ terms come from the contribution of $A_5$ which has an
effective bulk mass $m=1$.  $I_0(\Lambda)$ is roughly equal to the
number of KK modes with mass below $\Lambda T/k$.  It is defined
exactly in the previous section. There are additional terms in the
above equations proportional to $I_1(\Lambda) \frac{M_{GUT}}{k}$,
which we will assume to be small.  For $I_0=1$ (which occurs as
$\Lambda \ra 0$), these are just the standard equations for 4
dimensional running.  For bulk scalars, the effect is $n_s \ra n_s
I_0^{2,0}(\Lambda)$.  If the bulk preons are $n_f$ majorana fermions,
which would prevent them from picking up a large bulk mass, then this
term should be $n_s \ra 2n_f I_0^{1/2,0}(\Lambda)$, where $I_0^{1/2,
0}(\Lambda)$ comes from 5D massless fermion loops, and the factor of 2
is because fermions contribute twice as much as complex scalars to the
gauge boson self-energy. The same modifications should be made for the
$n_g$ term, but these do not affect unification, so we will ignore
them.

To show that unification can be improved, we pick a specific model. We
choose $\Lambda=1$ and put 4 majorana fermions in the bulk. We leave
$n_g=3$, to facilitate the comparison with the standard model. Then we
use the numerical values: $I_0(1) = 1.024$, $I_0^{1,1}(1)=0.013$,
$I_0^{1/2,0}(1) = 1.009$.  We will use the observed values \cite{lang}
of $\alpha_3(M_Z) = 0.1195$, $\alpha_e^{-1}(M_Z) = 127.934$, and
$\sin^2\theta_w = 0.23107$ at the Z-boson mass $M_Z = 91.187$ GeV.
The couplings are shown for this case in figure~\ref{unia5}. The
standard model is shown for comparison.

\FIGURE[t]{\centerline{\epsfig{file=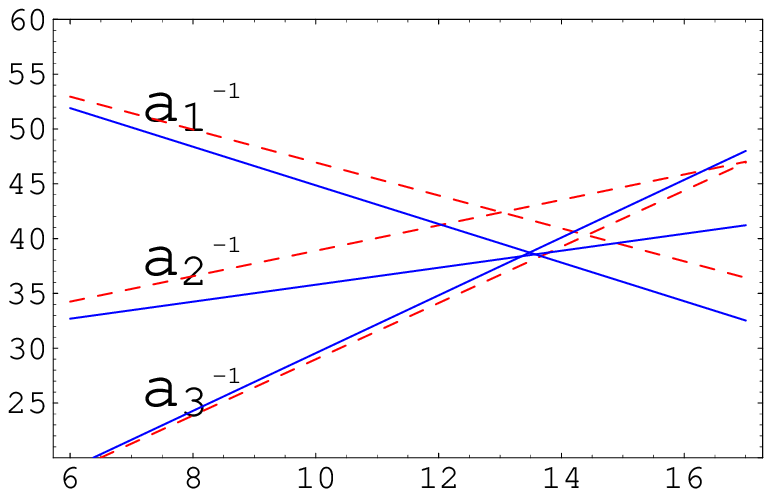,width=.5\textwidth}}%
\caption{$\alpha^{-1}$ as a function of $\log_{10} (M_{GUT}/M_Z)$.
Unification of couplings for $\Lambda = k$ (solid lines). The standard
model is shown for comparison (dashed lines).\label{unia5}}}

As $\Lambda$ is increased, $I_0(\Lambda)$ and $I_0^{1/2,0}(\Lambda)$
grow at roughly the same rate.  The net effect is that the
$\beta$-functions basically scale uniformly with $\Lambda$. This will
not have much of an effect on whether unification occurs, but it can
drastically change the scale, For example, if we take $\Lambda = 5$,
the scale drops from $10^{14}$ to $10^8$.  So, if we expect
unification near the string scale, we must have $\Lambda \approx 1$.
We assumed that $I(\Lambda,q)$ was constant. As we mentioned before,
the additional effect from the first order term, $I_1$ is suppressed
by $M_{GUT}/k$.  So if $M_{GUT} \ll k$ it is negligible, but if
$M_{GUT} \approx k$, it can be significant.  Even though our
regularization scheme cannot tell us the precise effect from the
1-loop calculations, we can easily determine the sign. $I_1(\Lambda)$
is the the slope of the curves in figure~\ref{31s}, and is always
negative.  So for $M_{GUT} \approx k$, these corrections will lower
the unification scale.

Now consider the second scenario, where the $XY$ bosons are decoupled
from the standard model by changing their $Z_2$ parity. Then the
coefficient of the log picks up an additional piece, proportional to
$I_0^d(\Lambda)$, as listed in the massless Dirichlet vector column of
table~\ref{tab.2}. Since complete multiplets do not contribute to unification,
we can simplify equations \eqref{uni1}--\eqref{uni3} by substituting:
\be
I_0(\Lambda) \ra I_0(\Lambda) -  I^d_0(\Lambda)\,.
\ee
The main effect of this is that it allows us to go to higher values of
$\Lambda$ without lowering the unification scale too much. For
example, $I_0(5) = 1.954$, but $I_0(5) - I^d_0(5) = 1.473$. This makes
$M_{GUT} \approx 10^{11}$ rather than $10^8$ as it would be without
these additional states.  We can also put in fields transforming as
adjoints or fundamentals under the GUT group with Dirichlet or Neumann
components. There are too many possibilities for us to examine them
here, but it is fairly straightforward to work out how they affect
unification.

Finally consider the third scenario, where matter is on the Planck
brane. Here $\SU(5)$ might be broken by a massive adjoint in the
standard way, and the triplet might be coupled to some heavy missing
partners. Proton decay is suppressed by at least $k^{-2}$, as we can
see from \eqref{xysup}.  Unification is similar to the second
scenario, but we must make the replacement $I_0(\Lambda) \ra
I_0(\Lambda) - I^{1,m}_0(\Lambda)$ in equations
\eqref{uni1}--\eqref{uni3}.  From table~\ref{tab.2}, we can see that
if the $XY$ bulk mass is $m=1$, the relevant value is $I_0(5) -
I^{1,1}_0(5) = 1.134$. This yields $M_{GUT} \approx 10^{13}$ even for
$\Lambda$ as big as $5k$.  However, if the bulk mass of $X$ and $Y$ is
too large, for example $m=5$, then $I_0(5) - I^{1,5}_0(5) = 1.776$,
which leads to $M_{GUT} \approx 10^{9}$.  It is clear that there is a
lot of room for detailed model building, which we leave for future
work.

\section{Conclusions}

We have shown how to consistently perform Feynman diagram calculations
in five-dimensional anti-de Sitter space.  Our regularization scheme
is inspired by AdS/CFT duality \cite{maldacena,witten,gubser}. There
we see that scale transformations in the 4D theory are equivalent to
z-translations in the 5D theory. Therefore, we can understand how
following the renormalization group flow down to the scale $\mu$
corresponds to integrating out the fifth dimension from $z=1/k$ up to
$z=\Lambda/(\mu k)$. The correct implementation of this is to renormalize
5D propagators as if the IR brane were at the relevant energy scale
for the computation. Not only does this ensure that at a position $z$
the UV cutoff is mediated by the warp factor, but also that a complete
4D mode of the bulk field is always present.

The original Randall-Sundrum scenario was presented as a solution to
the hierarchy problem. It is now clear that it is also consistent with
coupling constant unification. With standard model matter confined to
the TeV brane, the maximum unification scale is naively seen to be
$\Lambda T/k$. From the CFT picture, we know this cannot be true. Now
we understand how higher scales are reached in 5D as well. We have
briefly described some possible unification scenarios. None of them
are perfect, and more detailed model building is called for, but it is
clear that unification can be improved from the standard model. The
key is that even though gauge bosons are in the bulk, running is
effectively four dimensional.

Although we originally intended to tie up a loose end of the AdS/CFT
picture, it seems like we have revealed a whole new tangle. There are
many directions to go from here, and the work is to a large degree
unfinished. There are many threshold corrections that we have not yet
included. These include subleading terms in the background field
calculation, subleading terms in $I(\Lambda, q)$, and a higher loop
calculation.  Furthermore, the answer depends on the details of the
model; here it is not only a question of the GUT group, but also the
bulk fields that yield the TeV brane matter. It is well known that
while supersymmetric GUTs appear to unify beautifully at 1-loop, at
2-loops unification does not occur within experimental bounds (without
involved model building).  It is important to see in more detail how
well unification works in this model. Finally, we have been somewhat
lax about the relationships among the various scales in the theory,
namely $M_{GUT}, k, \Lambda k,M, M_{Pl}, R$ and the string scale.
These should ultimately be incorporated. Of course, we would also want
to motivate $\Lambda$ in a particular model, and furthermore
understand the origin of unification and its scale at a fundamental
level.

\acknowledgments

We would like to thank N.~Arkani-Hamed, A.~Karch, E.~Katz, M.~Porrati,
and F.~Wilczek for useful conversations. We especially thank N. Weiner
for helping to initiate this work.

\end{fmffile}

\end{document}